\documentstyle[12pt]{article}
\renewcommand{\thesection}{\arabic{section}}
\renewcommand{\theequation}{\thesection.\arabic{equation}}
\bibliography{plain}
\input{epsf}
\title{\bf Hierarchy of one and many-parameter families of elliptic chaotic maps of $\bf{cn}$ and $\bf{sn}$ types }
\author{M. A. Jafarizadeh$^{a,b,c}$\thanks{E-mail:jafarizadeh@tabrizu.ac.ir} and
   S. Behnia$^{b,e}$.\\
$^a${\small Department of Theoretical Physics and Astrophysics,
Tabriz University, Tabriz 51664, Iran.}
\\$^b${\small Institute for Studies in Theoretical Physics
 and Mathematics, Teheran 19395-1795, Iran.}
\\$^c${\small Research Institute for Fundamental Science,
 Tabriz 51664, Iran.}
\\ $^e $ {\small Department of Physics, IAU, Ourmia, Iran.}}
\begin{document}
\maketitle
\newpage
\begin{abstract}
We present hierarchy of one and many-parameter families of
elliptic chaotic maps of $\bf{cn}$ and $\bf{sn}$ types at the
interval $[0,1]$. It is proved that for small values of {\bf k}
the parameter of the elliptic function, these maps are
topologically conjugate to the maps of references
 \cite{jafar1,jafar2}, where using this  we have been able to
obtain the invariant measure of these maps for small {\bf k} and
thereof it is shown that these maps have the same Kolmogorov-Sinai
entropy or equivalently Lyapunov characteristic exponent of the
maps \cite{jafar1,jafar2}. As this parameter vanishes, the maps
are reduced to the maps presented in above-mentioned reference.
Also in contrary to the usual family of
 one-parameter maps, such as the logistic and tent maps, these maps do not
display period doubling or period-n-tupling cascade transition to
chaos, but they have single fixed point attractor at certain
parameter values where they bifurcate directly to chaos without
having period-n-tupling scenario exactly at these values of
parameters whose Lyapunov characteristic exponent begin to be
positive.
 {\bf PACs numbers}:05.45.Ra, 05.45.Jn, 05.45.Tp
\end{abstract}
\newpage
\section{Introduction}
In the past twenty years dynamical systems, particularly one
dimensional iterative maps have attracted much attention and have
become an important area of research activity \cite{Umeo1}
specially elliptic maps \cite{Cha, Umeo2}. Here in this paper we
propose a new hierarchy of families of one and  many-parameter
elliptic chaotic maps of the interval $[0,1]$. By replacing
trigonometric functions with Jacobian elliptic functions of ${\bf
cn}$ and ${\bf sn}$ types, we have generalized the maps presented
in references \cite{jafar1,jafar2} such that for small values of
{\bf k} the parameter of the elliptic function these maps are
topologically conjugate to the maps of references
\cite{jafar1,jafar2}, where using this  we have been able to
obtain the invariant measure of these maps for small {\bf k} and
thereof it is shown that these maps have the same Kolmogorov-Sinai
(KS) entropy  \cite{sinai} or equivalently Lyapunov characteristic
exponent of the trigonometric chaotic maps of references
\cite{jafar1,jafar2}. As this parameter vanishes, these maps are
reduced to trigonometric chaotic maps. Also it is shown that just
like the maps of references \cite{jafar1,jafar2}, the new
hierarchy of elliptic chaotic maps displays a very peculiar
property, that is, contrary to the usual maps, these maps do not
display period doubling or period-n-tupling, cascade transition to
chaos \cite{Ela} as their parameter $\alpha$ (parameters) varies,
but instead they have single fixed
 point attractor at certain region of parameters values, where they
bifurcate directly to chaos without having period-n-tupling
scenario exactly at the values of parameter whose Lyapunov
characteristic exponent begins to be positive.
\\ The paper is organized as follows: In section 2
we introduce new hierarchy of one-parameter families of elliptic
chaotic maps of $\bf{cn}$ and $\bf{sn}$ types, then in order to
make more general class of these families, by composing these
maps, we generate hierarchy of families of many-parameters
elliptic chaotic maps. For small values of the parameter of the
elliptic functions, we have presented, in section 3, the
equivalence of the elliptic chaotic maps with the trigonometric
chaotic maps of \cite{jafar1,jafar2}. In section 4 we obtain
Sinai-Rulle-Bowen measure for hierarchy of one and many-parameter
of elliptic chaotic maps for small ${\bf k}$. Section 5 is devoted to
explain KS-entropy of elliptic chaotic maps. Paper ends with a
brief conclusion.
\section{One-parameter and many-parameter families of elliptic chaotic maps of $\bf{cn}$ and $\bf{sn}$ types}
The families of one-parameter elliptic chaotic maps of $\bf{cn}$
and $\bf{sn}$ at the interval $[0,1]$ are defined as the ratio of
Jacobian elliptic functions of $\bf{cn}$ and $\bf{sn}$ types
\cite{wang} through the following equation:
 $$ \Phi_{N}^{(1)}(x,\alpha)
=\frac{\alpha^2\left(cn(N cn^{-1}(\sqrt{x}))\right)^{2}}{1+
(\alpha^2-1)\left(cn(N cn^{-1}(\sqrt{x}))\right)^{2}},$$
\begin{equation}
\Phi_{N}^{(2)}(x,\alpha) =\frac{\alpha^2\left(sn(N
sn^{-1}(\sqrt{x}))\right)^{2}} {1+(\alpha^2-1)\left(sn(N
sn^{-1}(\sqrt{x}))\right)^{2}}.
\end{equation}
Obviously, these equations map the unit interval $[0,1]$ into
itself. Defining shwarzian derivative \cite{dev}
$S\Phi^{\omega}_{N}(x), \omega=1,2$ as:
\begin{equation}
S\left(\Phi_N^{(\omega)}(x)\right)=\frac{\Phi_{N}^{(\omega)\prime\prime\prime}
(x)}{\Phi_{N}^{(\omega)\prime}(x)}-\frac{3}{2}\left({\frac{\Phi_{N}^{(\omega)
\prime\prime}(x)}{\Phi_{N}^{(\omega)\prime}(x)}}\right)^2=\left(
\frac{\Phi_{N}^{(\omega)\prime\prime}(x)}{\Phi_{N}^{(\omega)\prime}
(x)}\right)^{\prime}-\frac{1}{2}\left(\frac{\Phi_{N}^{(\omega)\prime\prime}
(x)}{\Phi_{N}^{(\omega)\prime}(x)}\right)^2,
\end{equation}
 with a prime denoting a single differential, one can show that:
$$ S\left(\Phi_{N}^{(\omega)}(x)\right)=S\left(sn(N
sn^{-1}(\sqrt{x})^{2}))\right)\leq0, $$
 since $\frac{d}{dx}\left(sn(N sn^{-1}(\sqrt{x})^{2}))\right)$ can be written
as: $$ \frac{d}{dx}\left(sn(N
sn^{-1}(\sqrt{x})^{2}))\right)=A\prod^{N-1}_{i=1}(x-x_i), $$
 with $0 \leq{x_1}<{x_2}<{x_3}<....<x_{N-1}\leq{1}$, then we
have: $$ S\left(sn(N
sn^{-1}(\sqrt{x})^{2}))\right)=\frac{-1}{2}\sum^{N-1}_{J=1}
\frac{1}{(x-x_j)^2}-\left(\sum^{N-1}_{J=1}\frac{1}{(x-x_j)}\right)^2<{0}.$$
Therefore, the maps $ \Phi_{N}^{(\omega)}(\alpha,x),$
 $\omega=1,2$ are (N-1)-nodal maps, that is, they have $(N-1)$
critical points in unit interval $[0,1]$ \cite{dev} and they have
only a single period one stable fixed point or they are ergodic
(See Figure 1).\\  As an example, we give below some of these maps
$\Phi_{2}^{(1)}(x,\alpha)$ and $\Phi_{2}^{(2)}(x,\alpha)$: $$
\Phi_{2}^{(1)}(x,\alpha)=\frac{\alpha^2\left((1-k^2)(2x-1)+k^2x^2\right)^2}
{\left(1-k^2+2k^2x-k^2x^2\right)^2+(\alpha^2-1)\left((1-k^2)
(2x-1)+k^2x^2\right)^2},$$
$$\Phi_{2}^{(2)}(x,\alpha)=\frac{4\alpha^2x(1-k^2x)(1-x)}{(1-k^2x^2)^2+
 4x(1-x)(\alpha^2-1)(1-k^2x)}.$$
Below we also introduce their conjugate or isomorphic maps which
can be very useful in derivation of their invariant measure and
calculation of their KS-entropy for small values of parameter {\bf k}.
Conjugacy means that the invertible map $ h(x)=\frac{1-x}{x} $
(which maps $ I=[0,1]$ into $ [0,\infty) $) transform maps
$\Phi_{N}^{(\omega)}(x,\alpha) $ into
$\tilde{\Phi}_{N}^{(\omega)}(x,\alpha), \omega=1,2$ defined as:
\begin{equation}
\left\{\begin{array}{l}
\tilde{\Phi}_{N}^{(1)}(x,\alpha)=\left(h\circ\Phi_{N}^{(1)}\circ
h^{-1}\right)(x,\alpha)=\frac{1}{\alpha^{1}}{\bf sc}^{2}(N {\bf
sc}^{-1}( \sqrt{x})),\\
\tilde{\Phi}_{N}^{(2)}(x,\alpha)=\left(h\circ\Phi_{N}^{(2)}\circ
h^{-1}\right)(x,\alpha)=\frac{2}{\alpha^{2}}{\bf cs}^{2}(N {\bf
cs}^{-1}(\sqrt{x})).
\end{array}\right.
\end{equation}
 Finally, by composing the maps introduced in $(2.1)$ we can define many-parameter
 families  of elliptic chaotic maps, where these many-parameter maps belong to
 different universal class than the single parameters ones (
  as it is shown at the end  of section 5). Therefore, by denoting their composition by
$\Phi_{N_{1},N_{2},\cdots,N_{n}}^{(\omega_1,\alpha_1),(\omega_2,\alpha_2),
\cdots,(\omega_n,\alpha_n)}(x)$ we can write:
\begin{eqnarray}
\nonumber
\Phi_{N_{1},N_{2},\cdots,N_{n}}^{(\omega_1,\alpha_1),(\omega_2,\alpha_2),
\cdots,(\omega_n,\alpha_n)}(x)=
\overbrace{\left(\Phi^{(\omega_1)}_{N_1}\circ\Phi^{(\omega_2)}_{N_2}\circ
\cdots\circ\Phi^{(\omega_n)}_{N_n}(x)\right)}^n=
\\
\Phi^{(\omega_1)}_{N_1}(\Phi^{(\omega_2)}_{N_2}(\cdots(\Phi^{(\omega_n)}_{N_n}
(x,\alpha_n),\alpha_{(n-1)})\cdots,\alpha_2),\alpha_1)
\end{eqnarray}
Thus obtained maps are many-parameter generalization of Ulam and
von Neumann maps \cite{von}. Since these maps consist of the
composition of the $(N_{k}-1)$-nodals $(N_{k}=1,2,\cdots,n)$ with
negative shwarzian derivative, they are $N_{1}N_{2}...
N_{n}-1$-nodals map and their shwarzian derivative is negative,
too \cite{dev}. Therefore, these maps have at most $N_{1}N_{2}...
N_{N}+1$ attracting periodic orbits \cite{dev}. Once again the
composition maps have also a single period one fixed point or
 they are ergodic (see Figure 2).\\
As an example, we give below some of them:
$\Phi_{2,2}^{(1,\alpha_{1}),(1,\alpha_{2})}(x)$ and $
\Phi_{2,2}^{(1,\alpha_{1}),(2,\alpha_{2})}(x)$:
 $$\Phi_{2,2}^{(1,\alpha_{1}),(1,\alpha_{2})}(x) =\frac{\alpha_1^2
\left(-{\bf Y}^2+2{\bf X}{\bf Y}+k_{1}^2 ({\bf Y}-{\bf
X})^2\right)^2} {\left({\bf Y}^2-k_{1}^2({\bf Y}-{\bf X}
)^2\right)^2+ (\alpha_1^2-1)\left(-{\bf Y}^2+2{\bf X{\bf
Y}}+k_{1}^2({\bf Y}- {\bf X} )^2\right )^2},$$ where:$$ {\bf
X}=\alpha_2^2\left(-1+2x+k_{1}^2(1-x
)^2\right)^2\;\;\mbox{and}\;\;
 {\bf Y}=\left (1-k_{1}^2(1-x )^2\right)^2+\left(\alpha_2^2-1)
(-1+2x+k_{1}^2(1-x)^2\right)^2$$ and $$
\Phi_{2,2}^{(1,\alpha_{1}),(2,\alpha_{2})}(x)
=\frac{\alpha_1^2\left(- {\bf Y}^2+2{\bf X}{\bf Y}+k_{1}^2( {\bf
Y}-{\bf X})^2\right)^2}{\left({\bf Y}^2-k_{1}^2( {\bf Y}-{\bf X}
)^2\right )^2 +(\alpha_1^2-1)\left(- {\bf Y}^2+2{\bf X}{\bf
Y}+k_{1}^2({\bf Y}-{\bf X} )^2\right)^2},$$where: $$ {\bf
X}=4\alpha_2^2x(1-x) (1-k_{2}^2x)\;\;\mbox{and}\;\;{\bf
Y}=(1-k_{2}^2x^2)^2+4(\alpha_2^2-1 )x(1-x) (1-k_{2}^2x).$$
\section{Topological conjugacy of elliptic chaotic maps with
trigonometric ones for small values of elliptic parameter ${\bf
K}$} \setcounter{equation}{0} In order  to obtain the SRB-measure
of one and many-parameter families of elliptic chaotic maps for
small value of elliptic parameter $\bf{k}$, we prove that elliptic
chaotic maps are topologically conjugated with trigonometric
chaotic maps of references \cite{jafar1,jafar2} for small value of
elliptic parameter.\\ To do so, we consider the first order
differential equation of elliptic chaotic maps given in $(2.1)$.
This differential equation can be
 obtained simply by taking derivation with respect to ${\bf x}$ from both sides of the
relations $(2.1)$, so we have:
\begin{eqnarray}
\nonumber\frac{d\tilde{\Phi}^{(\omega)}_{N}}{dx}=\frac{N}{\alpha}\times
\\ \nonumber \sqrt{\frac{\tilde{\Phi}^{(\omega)}_{N}
(1+\alpha^{2}\tilde{\Phi}^{(\omega)}_{N})(1+\alpha^{2}
\tilde{\Phi}^{(\omega)}_{N}-\frac{k^2}{2}((1-(-1)^{\omega})\alpha^{2}
\tilde{\Phi}^{(\omega)}_{N}+(1+(-1)^{(\omega)})))}
{x(1+x)(1+x-\frac{k^2}{2}((1-(-1)^{\omega})x+(1+(-1)^{\omega})))}}
\nonumber \\
\end{eqnarray}
 For small values of ${\bf k}$, the above differential equation is reduced to:
 \begin{equation}\frac{d
\tilde{\Phi}^{(\omega)}_{N}(x,\alpha)}{dx}= \frac{N}{\alpha}\times
\frac{ \sqrt{\tilde{\Phi}^{(\omega)}_{N}(x,\alpha)}
\left(1+\frac{(1-\frac{k^2}{4}(1-(-1)^{\omega}))
\alpha^{2}\tilde{\Phi}^{(\omega)}_{N}(x,\alpha)}
{1-\frac({k^2}{4}(1+(-1)^{\omega})}\right)}
{\sqrt{x}\left(1+\frac{(1-\frac{k^2}{4}(1-(-1)^{\omega}))
\alpha^{2}x}
{1-\frac{k^2}{4}(1+(-1)^{\omega})}\right)}.\end{equation} Now, the
dilatation map:
$$x^{\prime}=\frac{(1-\frac{k^2}{4}(1-(-1)^{\omega})) \alpha^{2}x}
{1-\frac{k^2}{4}(1+(-1)^{\omega})},\quad \tilde{\Phi}^{\prime\;
\omega}_{N}(x^{\prime},\alpha)=\frac{(1-\frac{k^2}{4}(1-(-1)^{\omega}))
\alpha^{2}\tilde{\Phi}^{\omega}_{N}(x,\alpha)}
{1-\frac{k^2}{4}(1+(-1)^{\omega})},$$ reduces the differential
equation $(3.1)$ to: $$ \frac{d
\tilde{\Phi}^{\prime\;(\omega)}_{N}(x^{\prime},\alpha)}{dx^{\prime}}=
\frac{N}{\alpha}
\frac{\sqrt{\tilde{\Phi^{\prime}}^{\omega}_{N}(x^{\prime},\alpha)}
\left(1+\alpha^{2}\tilde{\Phi^{\prime}}^{\omega}_{N}(x^{\prime},\alpha)
\right)}{\sqrt{x^{\prime}}(1+x^{\prime})}.$$ Integrating it, we
get:
\begin{equation} \tilde{\Phi}^{\prime\;(1)
}_{N}(x^{\prime},\alpha)=\frac{1}{(1-\frac{k^2}{2})}
\Phi^{(1)}_{N}((1-\frac{k^2}{2})x,\alpha)=\frac{1}{\alpha^{2}}{\bf
\tan}^{2}\left(N{\bf
\arctan}(\sqrt{x^{\prime}})\right),\end{equation}
\begin{equation} \tilde{\Phi}^{\prime\;(2)
}_{N}(x^{\prime},\alpha)=(1-\frac{k^2}{2})
\Phi^{(2)}_{N}(\frac{x}{1-\frac{k^2}{2}},\alpha)=\frac{1}{\alpha^{2}}{\bf
\cot}^{2}\left(N{\bf\arctan}(\sqrt{\frac{1}{x^{\prime}}})\right).\end{equation}
Therefore, for small values of ${\bf k}$ (the parameter of the
elliptic functions) elliptic chaotic maps are topologically
conjugate with trigonometric chaotic maps. Hence, for small ${\bf
k}$ their KS-entropy or equivalently Lyapunov characteristic
exponent is the same with the KS-entropy and Lyapunov exponent of
chaotic maps of reference \cite{jafar1,jafar2}, where the
numerical simulations of section 5 approve the above assertion.
Actually, the simulations of section 5 indicate that, except for
the values of ${\bf k}$ near one, the elliptic and trigonometric
maps are topologically conjugate. With a reasoning similar to one
given above, we can prove that the combination of elliptic maps
given in section 2 is
 almost topologically conjugate with the combination of trigonometric
maps of reference \cite{jafar2}.
\section{Invariant measure }
\setcounter{equation}{0} Characterizing invariant measure for
explicit nonlinear dynamical systems is a fundamental problem
which connects dynamical theory to statistics and statistical
mechanics. A well-known example is Ulam and von Neumann map which
has an ergodic measure $\mu=\frac{1}{\sqrt{x(1-x)}}$ \cite{von}.
The probability measure $\mu $  on $[0,1]$ is called an SRB or
invariant measure \cite{sinai}. For deterministic system such as
$\Phi_{N}^{(\omega)}(x,\alpha)$-map, the
$\Phi_N^{(\omega)}(x,\alpha)$-invariance means that, its invariant
measure $\mu(x)$ fulfills the following formal
Frobenius-Perron(FP) integral equation:
$$\mu(y)=\int_{0}^{1}\delta(y-\Phi_N^{(1,2)}(x,\alpha))\mu(x)dx.$$
This is equivalent to:
\begin{equation}
\mu(y)=\sum_{x\in\Phi_{N}^{-1(\omega)}(y,\alpha)}\mu(x)\frac{dx}{dy}\quad,
\end{equation}
defining the action of standard FP operator for the map
$\Phi_N(x)$ over a function as:
\begin{equation}
P_{\Phi_{N}^{(\omega)}}f(y)=\sum_{x\in
\Phi_{N}^{-1(\omega)}(y,\alpha)}f(x)\frac{dx}{dy}\quad.
\end{equation}
We see that, the invariant measure $\mu(x)$ is actually the
eigenstate of the FP operator $P_{\Phi_N^{(\omega)}}$
corresponding to the largest eigenvalue 1.\\ One can show that
$\tilde{\mu}(x)$, the invariant measure of conjugate map,
$\tilde{\Phi}=h\circ \Phi\circ h^{-1}$ can be written in terms of
$\mu(x)$, the invariant measure of chaotic map $\Phi$, as:
\begin{equation}
(\tilde{\mu}\circ h^{\prime})(x)=\mu(x)
\end{equation}
Therefore, considering the conjugacy relation (3.3) between the
maps $\Phi^1_{N}(x,\alpha)$ and\\
$\tilde{\Phi}_{N}^{1}(x,\alpha)=\frac{1}{\alpha^{2}}{\bf
\tan}^{2}\left(N{\bf \arctan}(\sqrt{x^{\prime}})\right)$ and using
the relation (4.3) with invertible map
$h(x)=\frac{1}{(1-\frac{k^2}{2})}$ together with taking into
account that the former one, $\Phi(x,\alpha)$, one has the
following invariant measure \cite{jafar1}
\begin{equation}
\mu(x, \beta)=\frac{1}{\pi}\frac{\sqrt{\beta}}{\sqrt{x(1-x)}
(\beta+(1-\beta)x)}\quad \quad \beta > 0,
\end{equation}
we obtain the following expression for the invariant measure of
chaotic maps $\Phi^1_{N}(x,\alpha)$:
\begin{equation}
\mu(x)=\frac{2\sqrt{2\beta}}{(2+(2+k^{2})\beta x)\sqrt{(2-k^2)x}},
\end{equation}
for small values of {\bf k}, where:
\begin{equation}
\alpha=\sqrt{\beta}\tan\left( {N
\arctan{(\sqrt{\frac{1}{\beta}})}}\right).
\end{equation}
With the same prescription as mentioned above, we can, for small
values of elliptic parameter {\bf k}, obtain the invariant measure
of all other types of elliptic chaotic maps which are
generalizations of trigonometric chaotic maps. It should be
mentioned that for trigonometric chaotic maps \cite{jafar1}, their
composition \cite{jafar2} and their coupling \cite{jafar3} the
invariant measure has already been obtained and presented in our
previous papers.
\section{KS-Entropy and Lyapounve exponent}
\setcounter{equation}{0}
KS-entropy or metric entropy measures how chaotic a dynamical system is and it
 is proportional to the rate at which information about the state of system is
lost in the course of time or iteration \cite{Dorf}. As it is
proved in Appendix A, for small values of {\bf k}, the KS-entropy
of elliptic  chaotic maps $h(\mu, \Phi^{\omega}_{N}(x,\alpha))$ is
equal to KS-entropy of trigonometric chaotic maps \cite{jafar1},
where for one-parameter elliptic chaotic maps we have:
\begin{equation}
 h(\mu,\Phi_{N}^{(\omega)}(x,\alpha))=\ln\left(\frac{N(1+\beta+2\sqrt{\beta})^{N-1}}{(\sum_{k=0}^{[ \frac{N}{2}]}C_{2k}^{N}\beta^{k})(\sum_{k=0}^{[
 \frac{N-1}{2}]}C_{2k+1}^{N}\beta^{k})}\right).
\end{equation}
\\ Also, in order to study discrete dynamical system, we could
refer to Lyapunov exponent which is, in fact, the characteristic
exponent of the rate of average magnificent of the neighborhood of
an arbitrary point $x_{0}$  and it is shown by $ \Lambda(x_{0}) $
which is written as: $$ \Lambda(x_{0})=lim_{n\rightarrow\infty}\ln
\left(\frac{d}{dx} \mid\overbrace{\Phi_{N}^{(\omega)}(x,\alpha)
\circ \Phi_{N}^{(w)}(x,\alpha) ....\circ
\Phi_{N}^{(\omega)}(x,\alpha)}\mid \right) $$
\begin{equation}
\Lambda(x_{0})=lim_{n\rightarrow\infty}\sum_{k=0}^{n-1}\ln\mid
\frac{d\Phi_{N}^{(\omega)}(x_{k},\alpha)}{dx}\mid,
\end{equation}
where $x_{k}=\overbrace{\Phi_{N} \circ \Phi_{N}\circ ....\circ
\Phi_{N}^{k}(x_{0})}$. It is obvious that $\Lambda(x_{0})<0 $ for
an attractor, $\Lambda(x_{0})>0$ for a repeller and
$\Lambda(x_{0})=0$ for marginal situation \cite{Dorf}. Also, the
Lyapunov number is independent of initial point $x_{0}$, provided
that the motion inside the invariant manifold is ergodic, thus
$\Lambda(x_{0})$ characterizes the invariant manifold of
$\Phi^{(\omega)}_{N}$ as a whole.\\ For small values of the
elliptic parameter, the map $\Phi^{(\omega)}_{N}$ and its
combination are measurable. Birkohf ergodic theorem implies the
equality of KS-entropy and Lyapunov
 number, that is \cite{Dorf}:
\begin{equation}
h(\mu,\Phi_{N}^{(\omega)}(x,\alpha))=\Lambda(x_{0},\Phi_{N}^{(\omega)}(x,\alpha)).
\end{equation}
A comparison of analytically calculated KS-entropy of maps
$\Phi_{N}^{(\omega)}(x,\alpha)$  $(5.1)$ and their combinations (
first kind ) for small values of the elliptic parameter, with the
corresponding Lyapunov characteristic exponent obtained by
simulation (see Figures $1-3$ ), indicates that in chaotic region
these maps are ergodic as Birkohf ergodic theorem predicts. In
non-chaotic regions of the parameters, Lyapunov characteristic
exponent is negative, since in this region we have only a single
period fixed point without transition to chaos.\\ Also, numerical
calculation shows that this class of maps have different
asymptotic behavior. Actually one can show that the KS-entropy of
one-parameter family of elliptic chaotic maps of ${\bf sn}$ and
${\bf cn}$ types $(5.1)$ have the following  asymptotic behavior:
\begin{equation}
\left\{ \begin{array}{l}
h(\mu,\Phi_{N}^{(\omega)}(x,\alpha=N+0^{-}))\sim
(N-\alpha)^{\frac{1}{2}}\\
h(\mu,\Phi_{N}^{(\omega)}(x,\alpha=\frac{1}{N}+0^{+}))\sim
(\alpha-\frac{1}{N})^{\frac{1}{2}},
\end{array}\right.
\end{equation}
Therefore, the above  relation $(5.4)$ implies that: all
one-parameter elliptic chaotic maps   belong to the same universal
class, which is different from the universality class of pitch
fork bifurcation maps, actually the  asymptotic behavior elliptic
ones  is similar to the class of intermittent maps \cite{pomeau}.
But intermittency can not occur in this family of maps for any
values
 of parameter $\alpha$ and for small values of parameter {\bf k} since elliptic
chaotic maps and their n-composition do not have minimum values other than zero
 and maximum values other than one in the interval $[0, 1]$.\\
It is interesting that the numerical and theoretical calculations
predict different asymptotic behavior for many-parameter elliptic
chaotic maps. As an example of asymptotic of the composed maps,
the KS-entropy of $\Phi^{(1,\alpha_1)(1,\alpha_2)}_{2,2}(x)$ is
presented below \cite{jafar2}:
\begin{equation}
h(\mu,\Phi^{(1,\alpha_1)(1,\alpha_2)}_{2,2}(x))=\ln\frac{(1+\sqrt{\beta})^2
(2\sqrt{\beta}+\alpha_2(1+\beta))^2}{(1+\beta)
(4\beta+\alpha_2^2(1+\beta)^2)},
\end{equation}
With choosing $\beta=\alpha_2^{\nu}\>, 0<\nu<2$, entropy given by
$(5.5)$ reads: $$
h(\mu,\Phi^{(1,\alpha_1)(1,\alpha_2)}_{2,2}(x))=\ln\frac{(1+\alpha_2^{\frac{\nu}{2}})^2
(2\alpha_2^{\frac{\nu}{2}}+\alpha_2(1+\alpha_2^{\nu}))^2}
{(1+\alpha_2^{\nu})(4\alpha_2^{\nu}+\alpha_2^2(1+\alpha_2^{\nu})^2)},
$$ which has the following asymptotic behavior near
$\alpha_2\longrightarrow 0$ and $\alpha_2\longrightarrow\infty$:
$$ \left\{ \begin{array}{l}
h(\mu,\Phi^{(1,\alpha_1)(1,\alpha_2)}_{2,2}(x))
\sim\alpha_2^{\frac{\nu}{2}}\quad \quad\mbox{as}\quad
\alpha_2\longrightarrow 0, \\
h(\mu,\Phi^{(1,\alpha_1)(1,\alpha_2)}_{2,2}(x))
\sim(\frac{1}{\alpha_2})^ {\frac{\nu}{2}}\quad\quad\mbox{as}\quad
\alpha_2\longrightarrow\infty.
\end{array}\right.$$
 The above asymptotic behaviours indicate that, for an  arbitrary value
of $0<\nu<2$, the maps $\Phi^{(1,\alpha_1)(1,\alpha_2)}_{2,2}(x)$
belong to the universal class which is different from that of
one-parameter elliptic chaotic maps $(5.4)$ or that of pitch fork
bifurcating maps.\\ In summary, combining the analytic discussion
of section 2 with the numerical simulation, we deduce that these
maps are ergodic in certain values of their parameters as
explained above and in the complementary interval of parameters
they have only a single period one attractive fixed point in a way
that, in contrary to the most of usual one-dimensional
one-parameter families of maps, they have only a transition to
chaos from a period one attractive fixed points to chaotic state
or vise versa.
\section{Conclusion}
 We have given hierarchy of one and many-parameter families of one-dimensional elliptic
 chaotic maps having the interesting property of being either chaotic (proper to say
ergodic ) or having stable period one fixed point and they go  to
chaotic  state from a stable single periodic state without having
usual period doubling or period-n-tupling scenario. Perhaps this
interesting property is again due to existence of invariant
measure for small values of the elliptic parameter.

\newpage
\renewcommand{\thesection}{A}
\renewcommand{\theequation}{\thesection-\arabic{equation}}
\setcounter{equation}{0}
 {\large \appendix{\bf Appendix A}}: KS-entropy of ellptic chaotic maps:
\\ In order to prove that KS-entropy for one and many-parameter
elliptic chaotic maps for small values of elliptic parameter would
be equal to KS-entropy of trigonometric chaotic maps \cite{jafar1,
jafar2}, the following statement should be considered taking into
account that $y=\Phi^{(\omega)}_{N}(x,\alpha)$:
$$\mu(x)dy=\sum_{x_{i}\in f^{-1}(y)}\mu(x_{i})dx_{i},$$
$$\tilde{y}=h(y)\quad \quad \tilde{x}=h(x)\quad \quad
\tilde{y}=\tilde{f}(\tilde{x}),$$ with $$\tilde{f}=h\circ f\circ
h^{-1},$$ $$(\tilde{\mu} \circ h^{\prime})(x)=\mu(x),$$
$$\tilde{h}\left(\mu,\Phi^{(\omega)}_{N}(x,\alpha)\right)=
\int{d\tilde{x}{\tilde{\mu}(x)\ln{\mid
(\frac{d\tilde{y}}{dx}\mid)}}}$$ $$ =\int{dx (\tilde{\mu}\circ
h)(x)h^{\prime}}(x)\ln{\left(
\frac{dy}{dx}\frac{h^{\prime}(y)}{h^{\prime}(x)}\right)}$$ $$=
\int{{dx\mu(x)\ln{(\frac{dy}{dx})}}}+\int{dx \mu(x)\ln{\left(
\frac{dy}{dx}\frac{h^{\prime}(y)}{h^{\prime}(x)}\right)}}$$
$$=h\left(\mu,\Phi^{\omega}_{N}(x,\alpha)\right).$$ since
$$\int{dx\mu(x)\ln{h^{\prime}(y)}}=\int{\ln{\left(\sum_{x_{i}inf^{-1}(y)}
\mu{(x_{i})dx_{i}}\right)}}=\int{dx\mu{(x)\ln{(h^{\prime}(y))}}}
=\int{dx\mu{(x)\ln{(h^{\prime}(x))}}}.$$ In the same way, one can
show that for small values of elliptic parameters, the KS-entropy
of  many-parameter families of elliptic chaotic maps  would be
equal to KS-entropy of many-parameter families of trigonometric
chaotic maps of Reference\cite{jafar2}
 \newpage
{\bf Figures Captions} \\
\\ Fig.1. The plot of Lyapunov exponent of $ \Phi_{2}^{(1)}(x,\alpha)$,
 versus the parameters $ \alpha $.
\\ Fig.2. The plot of Lyapunov exponent of $ \Phi_{2}^{(2)}(x,\alpha)$,
 versus the parameters $ \alpha $.
\\ Fig.3. The plot of Lyapunov exponent of $\Phi_{2,2}^{(1,\alpha_1),(1,\alpha_2)}(x)$,
versus the parameters $\alpha_1$ and $\alpha_2$.
\end{document}